\documentclass[nocopyrightspace]{sigplanconf}

\usepackage{lmodern}
\usepackage[utf8]{inputenc}
\usepackage[TS1,T1]{fontenc}
\usepackage[usenames,dvipsnames,svgnames,table]{xcolor}
\usepackage{url}
\usepackage{graphicx}
\DeclareGraphicsExtensions{.eps}
\usepackage{latexsym}
\usepackage{xspace}
\usepackage{enumitem}
\usepackage{subfigure}
\setenumerate{noitemsep,topsep=0pt,parsep=0pt,partopsep=0pt}

\newcommand{\idb}[1]{\ensuremath{\mathsf #1}} 	
\newcommand{\edb}[1]{{\ensuremath{\underline{\mathsf{#1}}}}}   
\newcommand*{\eat}[1]{}

\newcommand*{\wdl}{WebdamLog\xspace}

\newcommand*{\peer}[1]{\ensuremath{\mathsf{#1}}}
\newcommand*{\relName}[1]{\ensuremath{\mathsf{#1}}}
\newcommand*{\relduo}[2]{\relName{#1}\textbf{@}\peer{#2}}
\newcommand*{\reltri}[3]{\relName{#1}\textbf{@}\peer{#2}(\ensuremath{\mathsf{#3}})}

\newcommand*{\hide}{{{\sc hide}}\xspace}

\colorlet{comcol}{black!60}

\def \epf{alicePhotos}
\def \ipf{allPhotos}

\begin{document}

\title{Introducing Access Control in Webdamlog\thanks{This work has been
partially funded by the European Research Council under the European
Community's Seventh Framework Programme (FP7/2007-2013) / ERC
grantWebdam, agreement 226513. http://webdam.inria.fr/} }

\authorinfo{Serge~Abiteboul \and \'{E}milien~Antoine}
{INRIA Saclay \& ENS Cachan}
{first.last@inria.fr}

\authorinfo{Gerome~Miklau}
{INRIA Saclay \& UMass Amherst}
{miklau@cs.umass.edu}

\authorinfo{Julia~Stoyanovich \and Vera~Zaychik~Moffitt} {Drexel
University} {stoyanovich@drexel.edu \and zaychik@drexel.edu}

\date{\today}
\maketitle

\begin{abstract}
  We survey recent work on the specification of an access control
  mechanism in a collaborative environment. The work is presented in
  the context of the \wdl language, an extension of datalog to a
  distributed context.  We discuss a fine-grained access control
  mechanism for intentional data based on provenance as well as a
  control mechanism for delegation, i.e., for deploying rules at
  remote peers.
\end{abstract}

\section{Introduction}

The personal \emph{data} and favorite \emph{applications} of a Web
user are typically distributed across many heterogeneous devices and
systems, e.g., residing on a smartphone, laptop, tablet, TV box, or
managed by Facebook, Google, etc.  Additional data and computational
resources are also available to the user from relatives, friends,
colleagues, possibly via social network systems.  Web users are thus
increasingly at risk of having private data leak and in general of
losing control over their own information.  In this paper, we consider
a novel \emph{collaborative access control mechanism} that provides
users with the means to control access to their data by others and the
functioning of applications they run.

Our focus is information management in environments where both data
and programs are distributed.  In such settings, there are four
essential requirements for access control:
\begin{description}

\item [Data access] Users would like to control who can read and
  modify their information.

\item [Application control] Users would like to control which
  applications can run on their behalf, and what information these
  applications can access.

\item [Data dissemination] Users would like to control how pieces of
  information are transferred from one participant to another, and how
  they are combined, with the owner of each piece keeping some control
  over it.

\item [Declarativeness] The specification of the exchange of data,
  applications, and of access control policies should be
  declarative. The goal is to enable anyone to specify access control.

\end{description}

To illustrate each of these requirements, let us consider the
functionalities of a social network such as Facebook, in which users
interact by exchanging data and applications.  First, a user who wants
to control who sees her information, can use a classic access control
mechanism, such as the one currently employed by Facebook, based on
groups of friends.  Next, let us consider a user who installs an
application. This typically involves opening much of her data to a
server that is possibly managed by an unknown third party. Many
Facebook users see this as unreasonable, and would like to control
what the application can do \eat{``under their names''}on their
behalf, and what information the application can access. Third, with
respect to data dissemination, users would like to specify what other
users can do with their data, e.g., whether their friends are allowed
to show their pictures to their respective friends. Finally, the users
want to specify access control on their data without
having\eat{/knowing} to write programs. Thus, this simple example
already demonstrates the need for each one of the four above
requirements.

From a formal point of view, we define an access control mechanism for
\wdl, a declarative \emph{datalog}-style language that emphasizes
cooperation between autonomous peers~\cite{webdamlog}.  We obtain a
language that allows for declaratively specifying both data exchange
and access control policies governing this exchange.  There are
different aspects to our access control:

\begin{itemize}

\item For extensional data, the mechanism is standard, based on
  access control lists at each peer, specifying who owns and who can
  read/write data in each relation of that peer.

\item For intentional data, the mechanism is more sophisticated and
  fine-grained.  It is based on \emph{provenance}. In brief, only
  users with read access to all the tuples that participated in the
  derivation of a fact can read this fact.

\item The previous two mechanisms are used by default, and we also
  support the means of overriding them.

\item Finally, we introduce a mechanism for controlling the use of
  \emph{delegation} in \wdl, which allows peers to delegate work to
  remote peers by installing rules, and is one of the main
  originalities of \wdl.
\end{itemize}

Note that access control is implemented natively as part of the \wdl
framework.  The main idea is to use extensional relations to specify
access to extensional data. Thus accessibility of extensional facts is itself recorded as extensional facts, which can then be used by a \wdl
program to \emph{derive access} to intentional data.

\paragraph*{Organization} This short paper is organized as follows.
The Webdamlog language is presented in Section \ref{sec:webdamlog}. In
Section \ref{sec:access}, we present some aspects of access control.
We conclude in Section~\ref{sec:conc}.

\section{Webdamlog}
 \label{sec:webdamlog}

 In~\cite{webdamlog}, we introduced Webdamlog, a novel Datalog-style
 rule-based language.  In Webdamlog, each piece of information belongs
 to a {\em principal}.  We distinguish between two kinds of
 principals: {\em peer} and {\em virtual principal}.  A peer, e.g.,
 \peer{AlicePhone} or \peer{Picasa}, has storage and processing
 capabilities, and can receive and handle queries and update requests.
 A virtual principal, e.g., \peer{Alice} or \peer{RockClimbingClub},
 represents a user or a group of users, and relies on peers for
 storage and processing.  We further distinguish between {\em facts},
 representing local tuples and messages between peers, and {\em
   rules}, which may be evaluated locally or delegated to other peers.

Webdamlog is primarily meant to be used in a distributed
setting. Perhaps the main novelty of the language is the notion of
{\em delegation}, which amounts to a peer installing a rule on another
peer.  In its simplest form, delegation is a remote materialized view.
In its general form, it allows peers to exchange knowledge beyond
simple facts, providing the means for a peer to delegate work to other
peers.  We will not describe Webdamlog in detail here, but will
illustrate it with examples, referring the interested reader
to~\cite{webdamlog}.

  The following are examples of Webdamlog facts:

\eat{
\begin{tabbing}
~~{agenda}@{AlicePhone} (12/12/2012, 10:00, John, Orsay)\\
~~{photos}@{Picasa} (fileName:picture34.jpg, \\
~~~~~~~~~~~~~~~~~~~~~~~~date:09/12/2012, byteStream:010...001)\\
~~{writeSecret}@{Picasa} (login:Alice, password:HG-FT23)
\end{tabbing}
}
\begin{tabbing}
~~\reltri{agenda}{AlicePhone}{12/12/2012, 10:00, John, Orsay}\\
~~\relduo{photos}{Picasa}\ensuremath{\mathsf{(fileName:picture34.jpg,}} \\
~~~~~~~~~~~~~~~~~~~~~~~~\ensuremath{\mathsf{date:09/12/2012, byteStream:010001)}}\\
~~\reltri{writeSecret}{Picasa}{login:Alice, password:HG-FT23}
\end{tabbing}
The first fact represents a tuple in relation \peer{agenda} on peer
\peer{AlicePhone} with information about an upcoming meeting, and the
second, a photo in Alice's Picasa account (a tuple in relation
\peer{photos} on peer \peer{Picasa}). The third fact represents Alice's
login credentials for her Picasa account (in relation
\peer{writeSecret} on peer \peer{Picasa}).  Suppose that Alice wishes
to retrieve, and store on her laptop, photos from Fontainbleau outings
that were taken by other members of her rock climbing group.  To this
effect, Alice issues the following rule:

\eat{
\begin{tabbing}
{outingPhotos}@{AliceLaptop} ({\$pic}) :-\\
~~~{rockClimbingGroup}@{Facebook} ({\$member}),\\
~~~{findPhoto}@{AliceLaptop} ({\$member}, {\$photos}, {\$peer}),\\
~~~{\$photos}@{\$peer} ({\$pic}, {\$meta}),\\
~~~{contains}@{\$peer} ({\$meta}, ``Fontainbleau'')
\end{tabbing}
}
\begin{tabbing}
\reltri{outingPhotos}{AliceLaptop}{\$pic} :-\\
~~~\reltri{rockClimbingGroup}{Facebook}{\$member},\\
~~~\reltri{findPhoto}{AliceLaptop}{\$member, \$photos, \$peer},\\
~~~\reltri{\$photos}{\$peer}{\$pic, \$meta},\\
~~~\reltri{contains}{\$peer}{\$meta, Fontainbleau}
\end{tabbing}

This rule is a standard Webdamlog rule that illustrates various
salient features of the language.  First, the rule is declarative.
Second, the assignment of values to peer names (e.g., \peer{\$peer})
and relation names (e.g., \peer{\$photos}) is determined during rule
evaluation.  Third, for \peer{\$peer} assigned to a system other than
\peer{AliceLaptop} (e.g., \peer{Picasa} or \peer{Flickr}), the
activation of this rule will result in activating rules (by
delegation), or in some processing simulating them in other systems.
\eat{(e.g., Picasa or Flickr).}  The evaluation of rules such as this one is
performed by the Webdamlog system, which is responsible for handing
communication and security protocols, and also includes a datalog
evaluation engine, namely the Bud system\eat{ of Berkeley; see
}~\cite{Hellerstein10}.

The semantics of a Webdamlog rule depends on the location of the
relations occurring in this rule. Let \peer{p} be a particular
peer. We say that a rule is \emph{local} to \peer{p} if the relations
occurring in the body are all in \peer{p}; intuitively, \peer{p} can
run such a rule. The effect of a rule will also depend on whether the
relation in the head of the rule is local (to \peer{p}) or not and
whether it is extensional or intentional.

Generally speaking, Webdamlog supports the following kinds of rules.
\begin{itemize}

\item A. Local rule with local intentional head. These rules, like
  classical datalog rules, define local intentional relations,
  i.e., logical views.

\item B. Local rule with local extensional head. These rules derive
  new facts that are inserted into the local data\-base.  Note that,
  by default, as in Dedalus~\cite{dedalus}, facts are not
  persistent. To have them persist, we use rules of the form
  \reltri{m}{p}{U} \mbox{:-} \reltri{m}{p}{U}. Deletion can be
  captured by controlling the persistence of facts.

\item C. Local rule with non-local extensional head. Facts derived by
  such rules are sent to other peers and stored in an extensional
  relation at that peer, implementing a form of messaging.

\item D. Local rule with non-local intentional head.  Such a rule
  defines a new intentional relation at a remote peer based on local
  relations of the defining peer.

\item E. Non-local. Rules of this kind allow a peer to install a rule
  at a remote peer, which is itself defined in terms of relations of
  other remote peers.  This is the \emph{delegation} mechanism that
  enables the sharing of application logic by peers, for instance,
  obtaining logic (rules) from other sites, and deploying logic
  (rules) to other sites.

\end{itemize}

\section{Access control in Webdamlog}
 \label{sec:access}

We present three simple examples that highlight particular aspects of
access control in WebdamlLog.

\paragraph{Fine-grained access control on intentional data}

Suppose that an intentional relation \relduo{\idb{\ipf}}{Alice} has
been specified by Alice. Suppose that Alice gives the right to
friends, say Bob and Sue, to insert pictures into this relation.
Alice's friends can do this by defining the rules:
\begin{tabbing} \small
    {[}at~Bob] \= \reltri{\idb{\ipf}}{Alice}{\$f} \mbox{:-}
    \reltri{\edb{bobPhotos}}{Bob}{\$f}
    \\
    {[}at~Sue] \> \reltri{\idb{\ipf}}{Alice}{\$f} \mbox{:-}
    \reltri{\edb{suePhotos}}{Sue}{\$f}
\end{tabbing}
(Relation names \edb{bobPhotos} and \edb{suePhotos} are underlined to
indicate that they are extensional.)  \relduo{\idb{\ipf}}{Alice} is
intentional and is now defined as the union of
\relduo{\edb{bobPhotos}}{Bob} and \relduo{\edb{suePhotos}}{Sue}.

The {\sc read} privilege on \relduo{\idb{\ipf}}{Alice} is a
prerequisite to having access to the contents of this relation, but
access is also controlled by the provenance of each fact, making {\sc
  read} access fine-grained.  One can think of each intentional fact
as carrying its provenance, i.e., how it has been derived. In our
simple example of Alice's album, the provenance of a photo coming from
Bob will simply be the provenance token associated with the
corresponding fact at Bob.  Then, to be able to read a fact in
\relduo{\idb{\ipf}}{Alice} that is coming from
\relduo{\edb{bobPhotos}}{Bob}, Charlie will need {\sc read} access on
\relduo{\edb{bobPhotos}}{Bob}.  To see a slightly more complicated
example, suppose that a fact $F$ may be obtained by taking the join of
two base facts $F_1,F_2$; and that the same fact may be obtained
alternatively by projection of a fact $F_3$.  To access $F$ a peer
would need to have read access to its container (the relation that
contains it) as well as to facts that suffice to derive it, here $F_3$
or the pair $(F_1,F_2)$.  In other words, a peer that has {\sc read}
access to an intentional fact must have sufficient rights to derive
that fact.

\paragraph{Overriding the default semantics}

For intentional data, we use by default an access control based on the
full provenance of each fact. (If a fact is derived in several ways,
each derivation specifies a sufficient access right.) Access control
based on full provenance may be more restrictive than is needed in some
applications, and we provide the means to override it.  Consider the
following rule that Alice uses to publish her own photos to her
friends:
\begin{tabbing} \small
    {[}at~Alice] \= \reltri{\ipf}{\$x}{\$f} \mbox{:-}  \\
      \> \edb{\epf}@\peer{Alice}(\$f), [\hide~\reltri{friends}{Alice}{\$x}]
\end{tabbing}

Ignore the \hide annotation first.  This rule is copying the photos of
Alice's friends into their respective \idb{\ipf} relations.  A friend,
say Pete, will be allowed to see one of Alice's photos only if he is
entitled to read the relation \relduo{friends}{Alice}.  Now, it
may be the case that Alice does not want to share this relation with
Pete, and so Pete will not see her photos. The effect of the \hide
annotation is that the provenance of facts coming from
\relduo{friends}{Alice} is hidden.  With this annotation, Pete will be
able to see the photos. This feature is indispensable in preventing
access control from becoming too restrictive.

\paragraph{Controlling delegation}

Recall that general delegation allows rules with non-local relations
in the body.  This leads to significant flexibility for application
development and is the main distinguishing feature of the Webdamlog
framework.  It also creates challenges for access control.

The following example illustrates the danger of a simplistic semantics
for non-local rules.
Consider the two rules:
\begin{tabbing}
    {[}at~Bob] \= \relduo{\edb{message}}{Sue}(``I~hate~you\mbox{''}) \mbox{:-}
               \reltri{date}{Alice}{\$d}\\
               \> \reltri{\edb{aliceSecret}}{Bob}{\$x} \mbox{:-}  \\
      \> ~~~~~~\reltri{date}{Alice}{\$d},  \reltri{\edb{secret}}{Alice}{\$x}
\end{tabbing}
If we ignore access rights, by delegation, this results in running the
following two rules at Alice's peer:
\begin{tabbing}
    {[}at~Alice] \= \relduo{\edb{message}}{Sue}(``I~hate~you\mbox{''}) \mbox{:-}
                 \reltri{date}{Alice}{\$d}
\\
               \> \reltri{\edb{aliceSecret}}{Bob}{\$x} \mbox{:-}  \\
      \> ~~~~~~\reltri{date}{Alice}{\$d},  \reltri{\edb{secret}}{Alice}{\$x}
\end{tabbing}
Assuming \reltri{date}{Alice}{\$d} succeeds, then by the first rule
\peer{Alice} sends some hate mail to \peer{Sue}, and by the second it
sends the contents of the relation \relduo{\edb{secret}}{Alice} to
\peer{Bob}, even if \peer{Alice} did not give {\sc read} access on
this relation to \peer{Bob}.

The main reason for this problem is that (by the standard semantics of
Webdamlog) we are running the delegation rules as if they were run by
\peer{Alice}.  Under access control, we are going to run them in a
\emph{sandbox} with \peer{Bob}'s privileges.  So with the first rule, the
hate message will be sent but marked as coming from \peer{Bob}.  And with the
second, the data will be sent only if \peer{Bob} has {\sc read} access to
\relduo{\edb{secret}}{Alice}.  So, for a client $c$ delegating a rule to
a server, the semantics of delegation under access control policies
guarantees that:
\begin{itemize}

\item
If the rule has side effects (e.g., it results in the insertion of
tuples in the relation of another peer), the author of the update is
$c$.

\item
The access privileges with which the rule executes are those of $c$.
\end{itemize}
Note that, in practice, \peer{Alice} sends \peer{Sue} a message saying
that the author of the message is \peer{Bob}. So, \peer{Sue} may
question this fact and asks \peer{Alice} to prove that this is indeed
the case.  But if this is indeed the case, \peer{Alice} has the
delegation from \peer{Bob} to prove her good faith.

Delegation is at the heart of distributed processing. With delegation,
a peer \peer{p} can ask another peer \peer{q} to do some processing on
its behalf. A natural question is whether this will yield exactly the
same semantics (with possibly very different performance) as if
\peer{p} were getting the data locally and running a local
computation. It turns out that the semantics is different. This is
because \peer{q} will use data that (i) \peer{q} has access to; and
(ii) \peer{p} has access to (because of the sandboxing). On the other
hand, a local computation at \peer{p} is limited by (ii) but not by
(i).

\section{Conclusion}
\label{sec:conc}

The \wdl language has been introduced in~\cite{webdamlog}.  The system
has been implemented and different aspects have been demonstrated in
conferences~\cite{webdamexchange:demo,AbiteboulAMST13}.  The access
control mechanism is currently being implemented.  The fine-grained
mechanism for intentional data raises various issues. In particular,
the materialization of intentional relations may generate lots of data
if performed naively.  This is the topic of on-going research.

\bibliographystyle{abbrv}
\bibliography{biblio}

\end{document}